\documentclass[onecol]{epl2}

\title{ An exact analytical solution for generalized growth models driven by a Markovian dichotomic noise}
\shorttitle{An exact analytical solution for generalized growth
models}

\author{G. Aquino\inst{1}\thanks{E-mail: \email{gaquino@imperial.ac.uk}}
 \and M. Bologna\inst{2,3}\thanks{E-mail: \email{mauroh69@libero.it}}
 \and H. Calisto\inst{4}\thanks{E-mail: \email{hcalisto@uta.cl}}}
\shortauthor{G. Aquino \etal}

\institute{ \inst{1}Imperial College, Faculty of Natural Sciences
Flowers Building G24, South Kensington, SW7 2AZ London, UK\\
\inst{2}Center for Nonlinear Science, University of North Texas,
P.O. Box 311427, Denton, Texas 76203-1427 \\
\inst{3} Instituto de Alta Investigaci\'{o}n, Universidad de
Tarapac\'{a}-Casilla 6-D Arica, Chile\\
\inst{4}Departamento de F\'{\i}sica Facultad de Ciencias
Universidad de Tarapac\'a, Casilla 7-D Arica }

 \pacs{2.50.Ey}{Stochastic processes}

\abstract{Logistic growth models are recurrent in biology,
epidemiology, market models, and neural and social networks. They
find important applications in many other fields including laser
modelling. In numerous realistic cases the growth rate undergoes
stochastic fluctuations and we consider a growth model with a
stochastic growth rate modelled via an asymmetric Markovian
dichotomic noise. We find an exact analytical solution for the
probability distribution providing a powerful tool with
applications ranging from biology to astrophysics and laser
physics. }

\begin{document}

\maketitle

% main text

\section{Introduction}

\bigskip

\bigskip

\label{sec1} In this letter we focus our attention on a growth
model that was first used to describe the statistical behavior of
a population of individual species; for example, human population
growth. This is an area of interest several centuries old and is
perhaps one of the oldest branches of biology to be studied
quantitatively. The first model for human population growth was
proposed by Malthus in 1798. In 1838 Verhulst~\cite{Ver} corrected
this first model taking into account both the limitation of the
growth of a population due to the competition between individuals,
and the limitation on the density of the population that the
environment can support. An example of this would be the
limitation on the amount of food that the system is capable of
producing. The proposed equation is known as the logistic
equation, $\dot{x}=a_{0}x (1-x)$.

There are many examples of assemblies that consist of a number of
elements that interact through cooperative or competitive
mechanisms. Some important examples include species that share a
given environment, such as animals that live in the seas and
rivers of our planet; the components of the central nervous system
of a living being; transmission of diseases caused by different
types of viruses; interacting vortices in a turbulent fluid;
coupled reactions between different chemical elements that make up
our atmosphere; interactions between galaxies; competition between
different political parties; business companies; and negotiating
treaties of exchange between different countries.

In the last few decades generalizations of the Malthus-Verhulst
model have been applied to lasers physics~\cite{Lamb,Pariser} and
have been widely considered in scientific literature. In Ref.
~\cite{Zyg1}, the reader can find a large list of references that
use this type of model for many other physical processes such as
the saturation growth process of a population~\cite
{Horsthemke,Kampen,Haken,Goel0}, which is considered one of the
most successful models in the field of population dynamics. In
addition the Malthus-Verhulst model has found applications the
social sciences~\cite{Herman,Montroll}, autocatalytic chemical
reactions~\cite{Gardiner,Bouche}, biological and biochemical
systems~\cite{Haken}, grain growth in polycrystalline materials
~\cite{Pande}, cell growth in foam~\cite{Worner}, and as an
effective model for the description of the populations of photons
in a single mode laser~\cite{Haken,Sargent,Ogata,Arecchi}. The
Lotka-Volterra model that was introduced early last century
~\cite{Goel} is a useful model to describe the interaction between
two species.

Mathematical models can be constructed either intuitively or from
first principles to describe the phenomenon of competition and
cooperation of many of the afore mentioned assemblies. This leads
one to propose balance equations generally coupled and nonlinear.
They contain some parameters that must be determined empirically
or calculated from auxiliary equations. When the number of
interacting variables is large, the number of balance equations is
too large and therefore very difficult to solve. An example of
this is classical mechanics applied to many-body systems. One does
not generally know all the initial conditions and therefore it is
necessary to develop statistical methods for multiple coupled rate
equations that describe the behavior of the system far from
equilibrium. Some important aspects of any assembly of elements
that can be studied using statistical methods is its inherent
stability, that is its stability with respect to small changes in
growth rate and the introduction of new elements.

\section{The generalized Malthus-Verhulst model}

Our starting point is the generic stochastic differential equation
for a growth model driven by a Markovian dichotomic noise

\begin{equation}
\dot{x}=\left[ a_{0}(t)+a_{1}\xi (t)\right] x\frac{(1-x^{\mu })}{\mu }
\label{equat1}
\end{equation}%
where the deterministic growth rate $a_{0}(t)$ is perturbed by the
Markovian dichotomic noise $\xi (t)$, in which $a_{1}$ and $\mu $,
with $\mu\geq 0$, are free parameters. Our main objective is to
calculate the exact probability distribution for this generic
model.

The state space of the noise $\xi (t)$ consists only of two
levels, $ \left( \Delta _{1},-\Delta _{2}\right) $. This noise is
called asymmetric Markovian dichotomic noise and it is also known
as random telegraph noise. The temporal evolution of the
conditional probability $ P(\xi ,t\mid \xi _{0},t_{0})$ that
completely characterizes the process is given by the following
master equation~\cite{Horsthemke, Gardiner}

\begin{equation}
\frac{d}{dt}\left(
\begin{array}{c}
P(\Delta _{1},t\mid \xi _{0},t_{0}) \\
P(-\Delta _{2},t\mid \xi _{0},t_{0})%
\end{array}%
\right) =\left(
\begin{array}{cc}
-\lambda _{1} & \ \ \lambda _{2} \\
\ \ \lambda _{1} & -\lambda _{2}%
\end{array}%
\right) \left(
\begin{array}{c}
P(\Delta _{1},t\mid \xi _{0},t_{0}) \\
P(-\Delta _{2},t\mid \xi _{0},t_{0})%
\end{array}%
\right)  \label{equat2}
\end{equation}
where \ $\lambda _{1}$ and $\lambda _{2}$\ are the probabilities by unit
time of switching between states $\Delta _{1}$ and $\Delta _{2}$, so that $%
\tau _{j}=1/\lambda _{j}$ are the mean sojourn times in these
states. The stationary solution of Eq.~(\ref{equat2}) can be
obtained by setting

\begin{equation}
P(\xi ,\infty \mid \xi _{0},t_{0})=\frac{1}{\gamma }\left( \lambda
_{2}\delta _{\Delta _{1},\xi }+\lambda _{1}\delta _{-\Delta _{2},\xi }\right)
\label{equat3}
\end{equation}%
where $\gamma =\lambda _{1}+\lambda _{2}$. If the Markovian
dichotomic noise has Eq.~(\ref{equat3}) as the initial condition,
then $\xi (t)$ is a stationary process. From Eq.~(\ref{equat2}) it
follows that the mean value is

\begin{equation}
\left\langle \xi (t)\right\rangle =\frac{\lambda _{2}\Delta _{1}-\lambda
_{1}\Delta _{2}}{\gamma }.  \label{equat4}
\end{equation}
For the sake of simplicity, we require that the mean value $
\left\langle \xi (t)\right\rangle $ vanish. This means that

\begin{equation}
\lambda _{2}\Delta _{1}=\lambda _{1}\Delta _{2}=\omega _{0}.  \label{equat5}
\end{equation}
The correlation function is

\begin{equation}
\left\langle \xi (t)\,\xi (t^{\prime })\right\rangle =\frac{\lambda
_{1}\lambda _{2}}{\gamma ^{2}}\left( \Delta _{1}+\Delta _{2}\right) ^{2}%
\mathrm{e}^{-\gamma |t-t^{\prime }|}.  \label{equat6}
\end{equation}
Higher order correlation functions are more complicated. However,
since the correlation function given by Eq.~(\ref{equat6}) is
indistinguishable from the Ornstein-Uhlenbeck process, the
dichotomic noise found wide applications in building the models
~\cite{Horsthemke}. Furthermore, by an appropriate procedure of
limit the dichotomic noise converges at the Gaussian white noise
as the Ornstein-Uhlenbeck does, and it also converges at the white
shot noise~\cite{broeck}. Experimental evidences of the dichotomic
noise have been found frequently in the
litterature~\cite{a,b,c,Horsthemke}.

In Ref.~\cite{Zyg1} the author uses the direct method, which
consists of formally integrating the stochastic differential
equation and then taking the mean value over all realizations of
the stochastic process. This method allows analytical treatment of
the moments $\left\langle x^{n}(t)\right\rangle $ for different
types of noise, in particular for Gaussian white noise and  white
shot noise. In Ref.~\cite{Zyg2} the same author uses the inverse
Mellin transform to calculate the stationary probability
distribution. This last procedure appears to be difficult for two
reasons. The first reason is that the mathematical problem of
finding a distribution knowing its moments generally does not have
a unique solution. The literature refers to this as the classical
problem of moments~\cite{Widder}. The second reason is that indeed
it is a very hard task to find an analytical inverse Mellin
transform. For example if we consider the case with $\mu =1$ in
Eq.~(\ref{equat1}) and with $\xi (t)$ an Ornstein-Uhlenbeck
process, the moments can be expressed as the following integral
~\cite{Mannella}

\begin{equation}
\left\langle x^{n}\left( t\right) \right\rangle =\frac{1}{\sqrt{\pi }}%
\int_{-\infty }^{\infty }dz\frac{\mathrm{e}^{-z^{2}}}{\left[ 1+\left( \frac{%
1-x_{0}}{x_{0}}\right) \mathrm{e}^{-\left( a_{0}t+2a_{1}z\sqrt{t}\right) }%
\right] ^{n}}\ ,\qquad n=1,2,3,\ldots .  \label{equat7}
\end{equation}
To find the inverse Mellin transform the parameter $n$ has to be
considered a real parameter. This fact makes it very difficult to
perform the inversion.

Using a relatively simple procedure the authors in Ref.
~\cite{Calisto} provide the exact probability distribution for a
model like Eq.~(\ref{equat1}) when the noise $\xi (t)$ is given by
the Ornstein-Uhlenbeck process. A further generalization can be
found in Ref.~\cite{Spagnolo} where several cases of white
non-Gaussian noise are examined.

For $\mu =0$, Eq.~(\ref{equat1}) reduces to the Gompertz model
~\cite{Goel0}, $\dot{x} =\left[a_{0}(t)+a_{1}\xi (t)\right] x\ln x
$, and for $\mu =1$ it becomes  the logistic equation driven by
the dichotomic noise, namely

\begin{equation}
\dot{x}=\left( a_{0}(t)+a_{1}\xi (t)\right) x(1-x).  \label{equat8}
\end{equation}
Stochastic effects on Eq.~(\ref{equat8}) have been considered
frequently in the literature. Refs.~\cite{Horsthemke, Kampen,
Haken,Goel,Goel0} contain several applications and developments of
this model. In Refs.~\cite{Leung,Jackson,Mannella2,Mannella3} the
transient behavior has been investigated when the system is driven
by the same type of perturbation and the relaxation time of the
system is calculated as a function of noise intensity. In
Ref.~\cite{Mannella} the results of Ref.~\cite{Jackson} are
extended to the case in which $a_{0}$ is perturbed by a colored
Gaussian noise and confirmed by an analogical experiment, as well
as by numerical simulations. To analyze a cancer cell population
the authors of Ref.~\cite{Quan} consider the model
$\dot{x}=a\,x-b\,x^{2}+x\,\xi_{1}(t)+\xi _{2}(t)$ where $\xi
_{1}(t)$ and $\xi _{2}(t)$ are correlated Gaussian white noises.
They write the corresponding Fokker-Planck equation and analyze
the behavior of the stationary probability density.

\section{The exact probability distribution}

\label{sectwo}

The model described by Eq.~(\ref{equat1}) using the Stratonovich
approach, can be reduced to an elementary differential equation by
means of the transformation

\begin{equation}  \label{transform}
y=\ln \left( \frac{1-x^{\mu }}{\mu  x^{\mu }}\right)
\label{trans1}
\end{equation}
which leads to the equation
\begin{equation}  \label{simple}
\dot{y}=-a_{0}-a_{1}\xi (t).  \label{equat4_b}
\end{equation}
We emphasize that for the points $x=0$ and $x=1$, the
transformation Eq.~(\ref{transform}) does not hold. The behavior
of the system in these points has to be analyzed through a limit
procedure. Following Ref.~\cite{Sancho}, we can write the
stochastic Liouville equation for the density function $\rho
(y,t;\xi)$ of a set of realizations of Eq.~(\ref{equat4_b}) as
\begin{equation}
\frac{\partial }{\partial t}\rho(y,t;\xi )=a_{0}\frac{\partial
}{\partial y} \rho (y,t;\xi )+a_{1}\frac{\partial }{\partial y}\xi
(t)\rho (y,t;\xi ). \label{equat9}
\end{equation}%
Taking the mean value over all realizations of $\xi (t)$ we obtain

\begin{equation}
\frac{\partial p}{\partial t}=a_{0}\frac{\partial p}{\partial y}+a_{1}\frac{%
\partial p_{1}}{\partial y}  \label{equat10}
\end{equation}%
where $p\equiv p(y,t)=\left\langle \rho (y,t;\xi )\right\rangle $
and $ p_1\equiv p_{1}(y,t)=\left\langle \xi (t)\rho (y,t;\xi
)\right\rangle $. Next, using the well-known Shapiro-Loginov
formula for differentiation of exponentially correlated stochastic
functions~\cite{Shapiro}, we obtain the following differential
equation for the function $p_{1}(y,t)$

\begin{equation}
\frac{\partial p_{1}}{\partial t}\,=-\Lambda \,p_{1}+a_{0}\frac{\partial
p_{1}}{\partial y}+a_{1}\frac{\partial }{\partial y}\left\langle \xi
^{2}(t)\rho (y,t;\xi )\right\rangle  \label{equat11}
\end{equation}%
where $\Lambda = \lambda _{1}\ +\lambda _{2}$. Following Ref.
~\cite{Fulinski} we have $\xi ^{2}(t)=\Delta ^{2}+\Delta _{0}\xi
(t),$ where $\Delta ^{2}=\Delta _{1}\Delta _{2}$ and $\Delta
_{0}=\Delta _{1}-\Delta _{2}$, transforms Eq.~(\ref{equat11}) into
\begin{equation}
\frac{\partial p_{1}}{\partial t}\,=-\Lambda \,p_{1}+\left(
a_{0}+a_{1}\Delta _{0}\right) \frac{\partial p_{1}}{\partial
y}+a_{1}\Delta ^{2}\frac{\partial p}{\partial y}.  \label{equat12}
\end{equation}%
Taking the time derivative of Eq.~(\ref{equat10}) and combining it with Eq.~(%
\ref{equat12}) we finally obtain

\begin{equation}
\frac{\partial ^{2}\,p}{\partial t^{2}}\,-\left( 2a_{0}+a_{1}\Delta
_{0}\right) \frac{\partial ^{2}p}{\partial t\partial y}+\left(
a_{0}^{2}+a_{0}a_{1}\Delta _{0}-a_{1}^{2}\Delta ^{2}\right) \frac{\partial
^{2}p}{\partial y^{2}}-a_{0}\Lambda \frac{\partial p}{\partial y}+\Lambda
\frac{\partial p}{\partial t}=0 .  \label{equat13}
\end{equation}
The Shapiro-Loginov formula has a hypothesis stating the
statistical independence between $\xi (t)$ and $\rho(y,t;\xi )$ at
the initial time $t=0$. Therefore at the time $t=0$, we have
$p_{1}(y,0)=\left\langle \xi (t)\rho (y,t;\xi
)\right\rangle\mid_{t=0}=0$. Consequently the initial conditions
for Eq.~(\ref{equat13}) are

\begin{equation}
p(y,t)\!\!\mid_{t=0}=\delta\left( y-y_{0}\right),\,\,\,\,\, \frac{\partial }{%
\partial t} p(y,t)\!\!\mid_{t=0}=a_{0}\delta ^{\prime }\left( y-y_{0}\right).
\label{equat14}
\end{equation}
The following change of variables
\begin{equation}  \label{cv}
t =\tau ,\,\,\,\,\,\, y=z- a_{0}\tau
\end{equation}
further simplifies Eq.~(\ref{equat13}), so that we end up with

\begin{equation}
\left[\frac{\partial ^{2}}{\partial \tau^{2}} - a_{1}\Delta _{0}\frac{%
\partial ^{2}}{\partial \tau\partial z} -a_{1}^{2}\Delta ^{2}\frac{\partial
^{2}}{\partial z^{2}}+\Lambda \frac{\partial}{\partial \tau}\right]%
P(z,\tau)=0  \label{equat16}
\end{equation}
where $P(z,t)=p(y,t)$. A formal solution of Eq.~(\ref{equat16}),
satisfying the initial conditions (\ref{equat14}) and vanishing
for $z\to\pm\infty$, is given by

\begin{equation}
P(z,\tau)=e^{-\frac{\Lambda}{2}\tau}\int\limits_{-\infty}^{\infty} e^{\imath
k r} \left[\cos\frac{\lambda}{2}\tau+\frac{ \Lambda-\imath a_1 \Delta_0 k}{%
\lambda} \sin\frac{\lambda}{2}\tau\right] \frac{dk}{2\pi}  \label{app1}
\end{equation}
where

\begin{equation}
r=\left(\frac{a_1 \Delta_0}{2} \tau+z-z_0\right),\,\,\,\,\,\,\lambda= \sqrt{%
\left(\Delta_1+\Delta_2\right)^{2}a_1^{2}k^{2}+2\imath a_1\Delta_0\Lambda
k-\Lambda^{2} }. \label{def_r}
\end{equation}
Evaluating the integral we obtain

\begin{eqnarray}  \label{longsol}
P(z,\tau)&=&\frac{e^{-\frac{\Lambda}{2}\tau+u r}}{2
}\theta\left(v\tau- \mid\! r\!\!\mid\right)\left[\frac{\left(v\tau
+a_1\Delta_0 \frac{r}{ 2v} \right) b }{\sqrt{ v^{2}\tau^{2} -
r^{2}}}I_1\left( b\sqrt{ v^{2}\tau^{2} - r^{2}}\right) +\right.
\nonumber \\
&+&\left.\frac{\Lambda-a_1\Delta_0 u}{2v} I_0\left( b\sqrt{%
v^{2}\tau^{2} - r^{2}}\right)\right]+\nonumber\\
&+&\frac{e^{-\frac{\Lambda}{2}\tau+u
r}}{2}\left[\left(1+\frac{a_1\Delta_0}{ 2v}\right)
\delta\left(v\tau-r\right)+\left(1-\frac{ a_1\Delta_0}{2 v}
\right)\delta\left(v\tau+ r \right)\right],  \label{app2}
\end{eqnarray}
where $r$, as function of $z$, is defined by Eq.~(\ref{def_r}),
$I_n(q)$ are the modified Bessel's functions, $\theta(q)$ is the
step function, $\delta(q)$ is the Dirac's delta. Furthermore, we
defined in a compact manner
$$
u=\frac{a_1\Lambda \Delta_0 }{4 v^{2} },\,\,b=\frac{\sqrt{\Delta_1\Delta_2}%
\Lambda \mid\! a_1\!\!\mid}{2v^{2}},\,\,v=\mid\! a_1\!\!\mid \frac{%
\Delta_1+\Delta_2}{2}.
$$
With the help of Eqs.~(\ref{transform}) and (\ref{cv}) we may
write the solution as function of the original variables $x,t$ as

\begin{equation}  \label{px}
p(x,t)=\frac{\mu}{x\left(1-x^{\mu}\right)}P\left[ \ln \left(
\frac{1-x^{\mu }}{\mu x^{\mu }}\right)+a_0t,t\right].
\end{equation}

\section{Numerical Analysis}

\begin{figure}[ht]
\vspace{0.5cm} \center
\includegraphics[height=5 cm,
width=6.4cm]{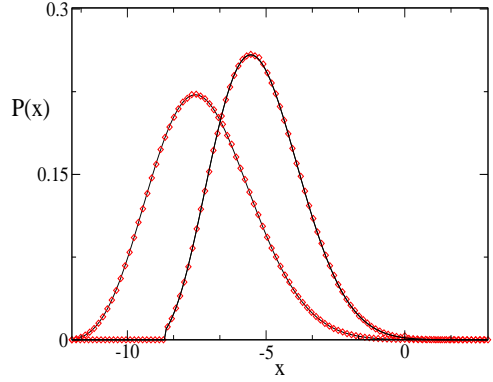} \vspace{0.3cm} \caption{The
probability distribution at times $t=5.7$ (right curve) and
$t=7.7$ (left curve) with parameters
$\protect\lambda_1=1.2,\protect\lambda _2=2.8, a_0=1.0, a_1=1.0,
\protect\mu=1.0, \Delta_1=0.6 ,\Delta_2=1.4$ . The diamonds are
the result of the numerical simulation, the continuous curves are
the analytic solutions as given by Eqs.~(\protect\ref{app2}) and
(\protect\ref{px}). The dirac deltas have been subtracted in both
cases.} \label{figa}
\end{figure}

In the following we numerically implement the dynamic equation
(\ref{simple}) driving the diffusion of the variable $y$. An
ensemble of N "trajectories" of the variable $\xi(t)$, the
dichotomic noise in (\ref{simple}), is produced via a generator of
random numbers with poisson distribution which returns the time
intervals in which the stochastic dichotomous variable $\xi(t)$
retains either of its two values. For each trajectory of $\xi(t)$,
a trajectory of the variable $y$ is obtained, integrating
Eq.~(\ref{simple}). The probability for a given value of $y$ at
time $t$ is then calculated as a simple average over the ensemble
of $N$ trajectories so obtained. A subsequent conversion to the
$x$-space through the transformation (\ref{transform}) allows us
to obtain the probability density $P(x,t)$. The dichotomous
process $\xi(t)$ is assumed in a stationary condition at time
$t=0$, i.e. at such a time $N \lambda_1/(\lambda_1+\lambda_2)$
trajectories are taken with initial value $\Delta_1$ for the
variable $\xi(t)$ and $N \lambda_2/(\lambda_1+\lambda_2)$ with the
value $-\Delta_2$. The numerical results are compared to the
analytical expression given by Eqs.~(\ref{app2}) and (\ref{px}) in
Fig.~(\ref{figa}) which shows a perfect agreement.

\section{Analysis of the results}\label{secthree}

Thanks to calculations performed in the previous sections we have
at our disposal the exact solution for any value of the parameter
$\mu$ of physical interest. A detailed study of $P(x,t)$ is beyond
the purpose of this letter. We shall limit ourselves to rediscover
some results known in the literature for the case $\mu=1$, which
is the Malthus-Verhulst model, and for $\mu=0$, which is the
Gompertz model.

We note a first result of solution~(\ref{equat16}), specifically
of Eq.~(\ref{app2}). Due to the asymmetric case under
consideration, there exists a particular choice of parameters such
that a coefficient of the two deltas vanishes. To fix the ideas
let us set $a_1>0$, then we can make one of the two delta
coefficients vanish if $a_1\Delta_0=2v$, that is to say
$\Delta_2=0$. Note that $\Delta_2=0$ does not imply a vanishing
value of the parameter $v$ that represents the propagation speed
of the peaks.

Taking appropriate limits on the parameters $\Lambda$, $\Delta_1$, and $%
\Delta_2$ we can rediscover different well known stochastic
processes. Following Ref.~\cite{broeck}, we consider the limit
$$
\lambda_1=\lambda_2= \lambda\to\infty,\,\,\,\,\Delta_1=\Delta_2=
\Delta\to\infty
$$
which corresponds to the Gaussian white noise. Keeping constant
the ratio $\Delta^2/\lambda$ we obtain

\begin{eqnarray}
P(z,\tau)&\approx&\frac{e^{-\lambda\tau}}{2}\theta\left(v\tau-
\mid\!
r\!\!\mid\right)\left[bI_1\left( b v\tau - \frac{b r^{2}}{2v\tau }\right) +%
\frac{ \lambda}{v} I_0\left(b v\tau - \frac{b r^{2}}{2v\tau
}\right)\right]
\nonumber \\
&+&\frac{e^{-\lambda\tau}}{2}\left[\left(1+\frac{a_1\Delta_0}{2v}\right)
\delta\left(v\tau-r\right)+\left(1-\frac{ a_1\Delta_0}{2 v}%
\right)\delta\left(v\tau+ r \right)\right].  \label{app3}
\end{eqnarray}
Using the asymptotic expression for the Bessel's functions that is
$I_n(x)\approx \exp[x]/\sqrt{2\pi x}$ for $x\to\infty$, we finally
obtain

\begin{equation}
P(z,\tau) \approx\frac{1}{\sqrt{2\pi D\tau}}\exp\left[-\frac{r^2}{2 D\tau}%
\right]  \label{app4}
\end{equation}
where by definition $D=a^{2}_{1}\Delta^{2}/\lambda$ and we neglect
the two exponentially damped deltas.

Still following Ref.~\cite{broeck}, to obtain the white shot noise
limit we have to take the symmetric dichotomous noise limit
$$
\lambda_1=\lambda_2= \lambda\to\infty,\,\,\,\,\Delta_1=\Delta_2=
\Delta\to\infty,\,\,\,\,\frac{\Delta}{\lambda}=\gamma
$$
where the $\gamma$ parameter is called non-Gaussianity parameter.
The relation with the diffusion coefficient $D$ is given by $
D=\gamma^2\lambda$. In the above limit $bv=\lambda\to\infty$ so
that we again end up with Eq.~(\ref{app3}) because the argument of
Bessel's functions becomes infinite. Consequently we rediscover
Eq.~(\ref{app4})

\begin{figure}[ht]
\vspace{0.5cm} \center\includegraphics[height=5 cm,
width=6.4cm]{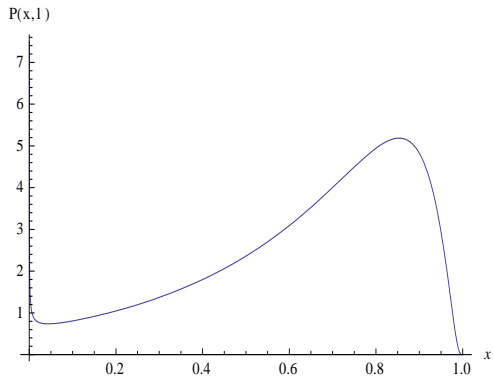} \vspace{0.3cm} \caption{The probability
distribution $P(x,t)$ for the Gompertz model at times $t=1$. The
value of the parameters are $a_0=1$ and $D=1$.} \label{figa2}
\end{figure}

As the last result we consider the limit for $\mu\to 0$ which
leads us to the Gompertz model. The transformation
(\ref{transform}) becomes

\begin{equation}\label{gomp}
y=\ln[-\ln x].
\end{equation}
For brevity we consider the symmetric case so that $\Delta_0=0$.
Using Eq.~(\ref{px}) we find that the asymptotic solution is

\begin{equation}
P(x,t) \approx-\frac{1}{\sqrt{2\pi D t}}\frac{1}{x \ln
x}\exp\left[-\frac{\left(\ln[-\ln x]+a_0 t\right)^2}{2D t}\right].
\label{gomp2}
\end{equation}
An analysis of this solution shows that $P(x,t)$, given by
Eq.~(\ref{gomp2}), always has a minimum located near the origin
and a maximum located near $x=1$. Finally $P(x,t)$ diverges at
$x=0$ as
$$P(x,t)\sim \frac{1}{x}\frac{1}{\left(-\ln
x\right)^{\frac{a_0}{D}+1}}.$$ These results are graphed in Fig.
\ref{figa2}.

\section{Concluding remarks}

In this letter we considered a growth model with a stochastic
growth rate modelled with a Markovian dichotomic noise. We found
an exact solution for the generic parameters $\mu$, $\Delta_1$ and
$\Delta_2$ which makes it possible to apply the result to a
variety of models ranging from biology~\cite{Goel0,Goel} to
astrophysics~\cite{chandra} and laser physics. For example our
model fits the the dichotomic noise condition found in ion channel
experiments~\cite{conti,ham, neher}. Patch-clamp experiments
detect a fluctuating dichotomous current which is related to the
number of ions flowing in the channel. Yet, in single mode lasers,
the growth model with $\mu=2$ and a constant growth rate,
describes the time evolution for the $N^{th}$ mode of the electric
field in the laser cavity~\cite{Lamb,Pariser}. The model in this
letter makes a case for dichotomic fluctuations around a resonance
frequency, providing opportunities for additional applications. An
interpretation of the Malthus-Verhulst model which describes the
photon population in a laser cavity, corresponding to the case
$\mu=1$, was introduced by McNeil and Walls~\cite{Walls}. We found
these to be interesting cases and chose to examine the
Malthus-Verhulst model, which has a very wide range of
applications such as dynamics of populations, chemical reactions,
and photon population in a laser cavity ~\cite{Bouche,Walls}. In
addition we studied the Gompertz model, corresponding to $\mu=0$,
when slightly modified, is used to describe the tumor growth
dynamics. We confirmed our results with a numerical simulation
showing perfect agreement with the analytical formulas. In this
letter we focused only on two models, but the obtained results may
applied to other systems that will be the subject of future
research.

\begin{acknowledgments}
M.B. thanks Welch (grant B-1577) for financial support of this
research work. H.C. thanks Derecci\'{o}n de Investigaci\'{o}n UTA
Project No 4720-08 for partial financial support. The authors are
indebted to Professor Paolo Grigolini for a critical reading of
the manuscript. The authors also thank Catherine Beeker for her
editorial contribution.
\end{acknowledgments}

%\section*{References}

\end{document}